\documentclass[hyper(*)]{JHEP3}

\usepackage{epsfig,multicol,bbm}
\usepackage{enumerate}
\usepackage{bibentry}

\title{Network, cluster coordinates and $N=2$ theory II: Irregular singularity}

\author{Dan Xie

\\ School of Natural Sciences, Institute for Advanced Study \\
Princeton, NJ 08540, USA}

\abstract {Cluster coordinates for a large class of  Argyres-Douglas  and asymptotical free theories are constructed using network on bordered Riemann surface. 
Such $\mathcal{N}=2$ theories are engineered using six dimensional $(2,0)$ theory on Riemann surface with irregular and regular singularities. The Stokes phenomenon 
plays an important role in our construction. Our results are expected to be very useful in studying BPS spectrum, wall crossing, and line operators of these theories, etc. 
In particular, we conjecture that the quiver from the network is the BPS quiver.  Moreover, our construction provides a
simple way to build the minimal network for cells of positive Grassmannia .}

\begin{document}

\section{Introduction}
A large class of four dimensional $\mathcal{N}=2$ Argyres-Douglas theory (AD) \cite{Argyres:1995jj,Argyres:1995wt} can be engineered from six dimensional $A_{k-1}$ $(2,0)$ theory using 
the irregular singularity \cite{Xie:2012hs}. Instead of specifying the UV Lagrangian theory and studying the scaling limit to find AD theory, our 
construction is much more simpler by just describing the form of the irregular singularity. Such construction is very powerful in 
answering many physical questions:  all the dimensional coupling constants and the 
number of mass parameters can be read directly from the form of irregular singularity; the scaling dimensions of chiral primary operators of the theory can be easily listed; 
the Seiberg-Witten curve and  three dimensional mirrors can be written down pretty easily; the central charges can be quickly
calculated; these AD theories can be used nicely in forming new asymptotical free theories which also have  six dimensional constructions, etc.

Many other dynamical properties of these theories can be found from this magical geometric construction:
the BPS spectrum and wall crossing behavior,
the classification of line operators and its expectation value could be extracted from the combinatoric
object defined on the Riemann surface.
The moduli space ${\cal M}$ of Hitchin equation with irregular singularity plays a central role here. ${\cal M}$  is the Coulomb branch
of the four dimensional theory compactified on a circle and is a Hyperkahler manifold \cite{MR2004129}.  A special coordinate system
called cluster coordinates \cite{Fomin2001} on framed \footnote{Framed can be thought of adding some extra parameters to the moduli space.} version of ${\cal M}$ seems to be really important in studying these properties as the study in $SU(2)$ case indicated: 

1.  Finding the BPS spectrum and studying the wall crossing behavior of a  4d $\mathcal{N}=2$ theory \cite{Gaiotto:2008cd,Gaiotto:2009hg,Cecotti:2011rv, Alim:2011ae, Alim:2011kw}.

2. The classification of the line operators \cite{Drukker:2009tz}, the BPS wall crossing in the presence of line defects and surface defects \cite{Gaiotto:2010be, Gaiotto:2011tf}.

Given the importance of the cluster coordinates, it would be really interesting to find such coordinates for the AD theory and asymptotical free theory constructed in \cite{Xie:2012hs}.
The purpose of this paper is to achieve this goal. The essential definition of the cluster coordinates involves a quiver, or more precisely a quiver mutation class. We 
are going to find quiver mutation class for these $\mathcal{N}=2$ theories using simple combinatorial methods.

${\cal M}$ in one of  complex structure describes the flat connections with irregular singularity \cite{MR2004129,Witten:2007td}. The solutions of the flat section equation 
have the very interesting Stokes phenomenon and  the moduli space can be defined using Stokes matrices. Basically, Stokes phenomenon
captures how the asymptotical behavior of the solutions changes in different angular region around the singularity. 
The Stokes matrix is actually a unipotent subgroup and one could use a Young Tableaux to label it, this is the most important starting point
for our construction. We blow up the irregular singularity into a disc and put a marked point with a Young Tableaux label for each Stokes matrix. 
Given such a cyclic ordering of the Young Tableaux, the method proposed in \cite{Xie:2012dw} can be used
to construct a dot diagram and  a bipartite network from which a quiver can be read.

The detailed procedure for finding the quiver is the following: Start with a triangulation of the disc with labeled marked points, and decorate the internal and 
external edges using black and white dots based on the information of the punctures. Such decoration is 
uniquely fixed by the Young Tableaux and the triangulation.  Each triangle is further tessellated 
using two types of minimal polygons and then a bipartite network could be found from the dot diagram. One only need to 
know the quiver for the triangle and the whole quiver is derived by gluing the quivers of the triangles together.

The first use of the quiver is to confirm some of the isomorphism proposed in \cite{Xie:2012hs}. The same theory can be realized by using different 
$A_N$ theory and different singularity combinations, we show that different configurations give the quiver in the same mutation class.
Our result agrees with what is found in the literature for those theories whose BPS quiver is known in some form. Our construction is very simple and purely combinatorial, moreover,
our result  has another virtue:
the quiver mutation sequences which leads back to the original quiver can be easily found, such sequences are very important in finding 
the BPS spectrum and studying the wall crossing behavior.  This result has great meaning since generically the quiver studied in this paper
is in a mutation infinite class, it is very hard to find the above sequences if we do random quiver mutations.
Moreover, our quiver and the cluster coordinates are actually describing a very interesting moduli space.

This paper is organized as follows: Section 2 review the basic data for the irregular singularity and regular singularity relevant for the 
AD theory constructed in \cite{Xie:2012hs}, we also review how to find the quiver based on a cyclic ordering of Young Tableaux. Section 3 describes 
the Stokes phenomenon and the Young Tableaux label for the irregular singularity. Section 4 is a detailed discussion on 
how to find the quivers for  the AD theory, many examples are given. Section 5 discusses the construction of the quiver for 
non-conformal theory. Finally, we give a short discussion on the possible applications of our result in section 6.

\section{Review}
\subsection{UV and IR data of AD theory}
A large class of four dimensional SCFTs can be engineered by compactifying six dimensional $A_{k-1}$ $(2,0)$  theory on 
a punctured Riemann surface. The theory of class ${\cal S}$ \cite{Gaiotto:2009we,Gaiotto:2009hg} is engineered by putting arbitrary number of 
regular singularities (first order pole) on a Riemann surface. Theory ${\cal S}$ has integer scaling dimensions and have 
dimensionless gauge coupling constants. On the other hand, the Argyres-Douglas (AD) theory has dimensional coupling constant and
irregular singularity (higher order pole) is needed.  Moreover, there are two constraints on the choice of the Riemann surface and the 
combination of the singularities: first, only Riemann sphere can be used; second, only one irregular singularity or one irregular singularity and one regular singularity at opposite poles are viable.

The regular singularities are classified by Young Tableaux which specifies a partition $k=n_1+n_2+...+n_r$ . We use the notation
$Y=[n_1,n_2,\ldots, n_r]$ to denote the Young Tableaux. There are $r-1$ mass parameters encoded in this puncture and the
corresponding flavor symmetry could also be read from $Y$.  There are two special punctures which we want to give them special name:
The full puncture has partition $[1,1,1..,1]$ whose flavor symmetry is $SU(k)$; The simple puncture has the partition $[k-1,1]$ and 
the flavor symmetry is $U(1)$.

The irregular singularity useful for the AD theory are much more fruitful and a complete classification is 
given in \cite{Xie:2012dw}:

a. Type I singularity: the holomorphic part of the Higgs field \footnote{ The Higgs field is a one form field defined on the Riemann surface, and there is also a gauge field appearing in Hitchin's equation.  
The Hitchin's moduli space is defined as the space of the solutions with the specified  boundary condition on these fields.}
\begin{equation}
\Phi={1\over z^{n+j/k+2}}diag(1,\omega,\omega^2,...\omega^{k-1})
\end{equation}
with $\omega=\exp({2\pi i\over k})$ and $z$ a local coordinate on Riemann sphere.  Here $n\geq -1$ is an integer and $0\leq j<k$. All the subleading terms compatible with the leading order form are allowed \footnote{A gauge transformation is needed to make the solution consistent around 
the singularity, we require the subleading singular terms should also be consistent with this gauge transformation, i.e. they are going back to  themselves after circling around the singularity followed by the gauge transformation.}. The AD 
theory for this class of irregular singularities are first studied in \cite{Cecotti:2010fi}.

b. Type II singularity: the Higgs field has the form
\begin{equation}
\Phi={1\over z^{n+j/(k-1)+2}}diag(0,1,\omega,\omega^2,...\omega^{k-2})
\end{equation}
with $\omega=\exp({2\pi i\over k-1})$ and $0< j<k-1$. We ignore the subleading terms compatible with the leading order behavior. Several lower rank examples of this class are discussed in \cite{Cecotti:2011gu,DelZotto:2011an}.

c. Type III singularity: The leading order pole is integer but the coefficient is not regular semi-simple. The classification is given by a 
sequence of Young Tableaux such that $Y_n\subset Y_{n-1} \subset \ldots \subset Y_1$. 

The first and second irregular singularity are nicely reflected geometrically by a Newton polygon: Starts with the unit two dimensional lattice and 
label the horizontal coordinate as $z$ and the vertical coordinate as $x$ \footnote{$x$ is the coordinate on cotangent bundle.}. The Newton polygon 
is formed by a sequence of lines ending on lattice points in the region $(x\geq 0,z\geq 0)$. 

The slop of  type I singularity is $r=n+j/k$, while the slops of two edges of type II singularities are $r=n+j/(k-1)$ and $r=n$ respectively. 
Many information of the four dimensional
SCFT can be read from the corresponding Newton polygon, i.e.  and the
number of mass parameters are the number of integer points on the newton polygon.
\begin{figure}[htbp]
\small
\centering
\includegraphics[width=10cm]{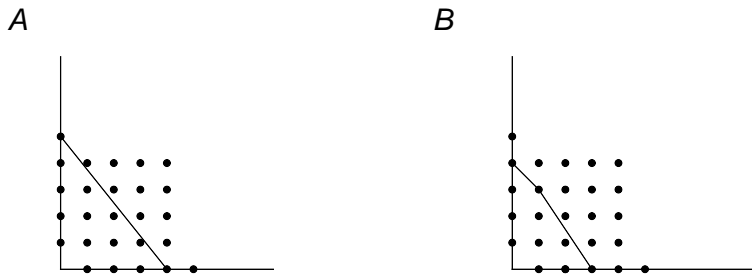}
\caption{The Newton polygon for type I and type II irregular singularity. }
\label{irregular}
\end{figure}

The IR physics are described by the moduli space of Hitchin's equation with above specified boundary conditions. 
In particular, the Seiberg-Witten curve is identified with the spectral curve of the Hitchin system. In the case of 
AD theory, one can easily find the full Seiberg-Witten curve from the corresponding Newton polygon of the irregular
singularity,, i.e. the monomials appearing in Seiberg-Witten curve 
is determined by the non-negative integer lattice points bounded by the Newton polygon. More details can be found in \cite{Xie:2012hs}.

The AD theory constructed in this way include almost all the AD theories found in the literature, and we have a huge number of 
new examples. Moreover, the theory defined using regular singularities on the sphere can all be realized using the irregular singularity.

\subsection{Dot diagram, network and quiver}
The result in \cite{Xie:2012dw} is crucial for our construction of quivers. This subsection serves as a light review of the relevant construction. 
Given a triangle labeled by three Young Tableaux, one could find a dot diagram and a further tessellation of the triangle using the brane construction proposed in \cite{Benini:2009gi}. 
 The three vertices of the triangle have coordinates $(N,0), (0,0), (0,N)$ in a two dimensional lattice with unit spacing.  The dot diagram is found as follows:
 
a. Decorate the boundary of the triangle using the information of the puncture, the white dots is used 
to represent the boxes on the same column of the Young Tableaux. 

b. Construct the dot diagram inside the triangle using only following two types of polygons whose edge is formed by lines \footnote{The lines should be parallel  with the boundary edges.} connecting two black dots. 1: Triangles whose edges have the same length. 2: Trapeziums whose 
parallel sides have lengths $n_1, n_2$ and the other two sides have length $n_1-n_2$ \footnote{This constraint is from the supersymmetric condition on the brane configuration.}.

\begin{figure}[htbp]
\small
\centering
\includegraphics[width=10cm]{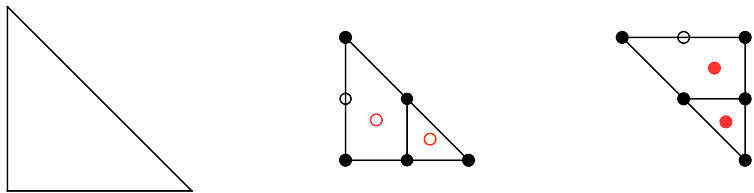}
\caption{Left: The orientation of the big triangle.  Middle: Put a white vertex to each polygon whose triangle completion has  the same orientation as 
the big triangle. Right: Put a black vertex to each polygon whose triangle completion has opposite orientation.}
\label{vertex}
\end{figure}

To state our rules for constructing the bipartite network \footnote{ A bipartite network has vertices colored with the black or white, and there are 
no edges connecting the vertices with the same color.}, we need to distinguish two types of polygons in the tessellation of the big triangle:
The type A polygon is the one whose triangle completion has the same orientation as the big triangle, and
the triangle completion of the type B polygon has opposite orientation.  The colored vertices in each polygon of the tessellation are determined in the following way (see figure. \ref{vertex}):

a: Assign a white vertex to each type A polygon. 

b: Assign a black vertex to each type B polygon.

 A network is formed by connecting the white vertex and black vertex if there is a common edge between two corresponding polygons. 
 We never connect two vertices with same color even if the corresponding polygons have one common edge.  Finally, there is one line  coming 
 out of boundary for the vertex inside the polygon which has one piece of boundary edge of the triangle. The network formed in this way is always bipartite but there may be vertices with only two edges. 
 We can use various moves to remove these degree-2 vertices and  get another bipartite network from which a quiver without two cycles can be read in the following two steps:
 
 Step1: Remove degree two vertices and then use the contraction to form another bipartite graph.
 
 Step 2:  Assign a quiver node to each surface and the quiver arrows are determined by the black vertices: there is 
 a clockwise closed circles around it.

\begin{figure}[htbp]   
\small
\centering
\includegraphics[width=4cm]{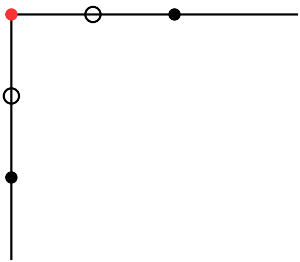}
\caption{The network would be non-minimal if the boundary of the dot diagram has this form at any vertex.}
\label{nonmini}
\end{figure}

 The three punctured network is a basic building block for more complicated cases. For more punctures, we start with a triangulation and 
 decorate the internal edges which is derived by using the gauge theory result \cite{Nanopoulos:2010ga}. After these decorations on the edges of the triangulation,
 we can do the tessellations on each triangle and find the network, quiver, etc.  Different triangulations are related by the a sequence of so-called flip which relate 
  two triangulations of the quadrilateral, we have proved that if the dot diagram on all the punctures does not have the  "bad" configuration shown in figure. \ref{nonmini}, the corresponding 
  quivers are related by quiver mutation (more precisely the two networks are related by square moves).  However, if we consider only the quiver nodes represented
  by the closed surfaces, even with the "bad" corner, the quivers from different triangulations can still be isomorphic as we show later.
  
The planar network has been studied in greater detail in describing the cells of the positive Grassmannia in \cite{postnikov-2006}. Our result in this paper
will give some simple ways of constructing minimal network.  The higher rank generalization of the cluster coordinates for boarded Riemann surface is given in a seminal
paper by Fock and Goncharov \cite{fock-2003}, and the similar network construction  is also recently considered by Goncharov \cite{Goncharov:2012aa}.

  \subsection{Quiver and mutation}
\begin{figure}[htbp]
\small
\centering
\includegraphics[width=10cm]{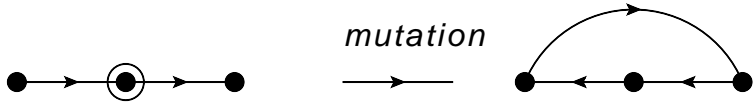}
\caption{The quiver mutation. }
\label{seiberg}
\end{figure}

 A quiver is a directed graph where multiple arrows between two vertices are allowed.  The quiver mutation for a quiver without one and two cycles is 
 defined as the following: Let Q be a quiver and k a vertex of Q. The mutation $\mu_k(Q)$ is the quiver obtained from Q as follows, see figure. \ref{seiberg}:

1) for each sub quiver $i\rightarrow k\rightarrow j$, create a new arrow between $ij$ starting from $i$;

2) we reverse all arrows with source or target k;

3) we remove the arrows in a maximal set of pairwise disjoint 2-cycles. 
 
\begin{figure}[htbp]
\small
\centering
\includegraphics[width=10cm]{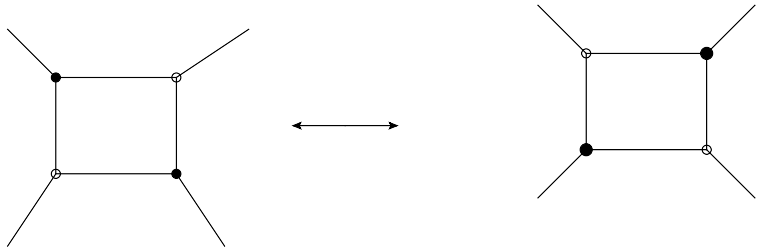}
\caption{The square move which represents the quiver mutation. }
\label{move}
\end{figure}

 Geometrically, a quiver could be found from the triangulation of the boarded Riemann surface and the quiver mutation is 
 realized as the flip on the triangulation. The quiver can also be found from a bipartite network using the rule stated in last subsection and the quiver mutation 
 is represented perfectly by the square move, see figure. \ref{move}. One can check the two quivers are related by quiver mutation. Notice that
 the above two geometrical quiver mutations are special since the node under mutation always has four quiver arrows on it.

 Quiver mutation class of a quiver is defined as a collection of all the quivers which could be got by doing quiver mutation on it. 
 Quiver with finite mutation class is the one with only finite number of quivers in this class.  
 Most of the finite mutation class are associated with the triangulated surface of $A_1$ theory \cite{fomin-2008-201}.  The $ADE$ 
 quiver is kind of special since the corresponding cluster coordinates are also finite.  $A$ type and $D$ type quivers can be found from the triangulated surface. 
 In this paper, we are going to show that the $E$ type quiver can also be found from the triangulated surface but with $A_2$ group.
 
 Quivers are used to study the BPS spectrum of $\mathcal{N}=2$ theory and the corresponding
 $\mathcal{N}=2$ theory for  finite quiver mutation classes are found in \cite{Cecotti:2011rv}. One of the motivation of this paper is 
 to find the quiver for other $N=2$ theory related to general Argyres-Douglas theory discussed in \cite{Xie:2012hs}, and most of 
 the quivers discovered in this paper are mutation infinite.

\section{Stokes phenomenon}
The Coulomb branch of four dimensional field theory compactified on a circle is described by 
the Hitchin's moduli space with irregular singularity. The purpose of this paper is to 
find the cluster coordinates for it.  In one of the complex structure of the Hitchin's moduli space,
each point describes a complex flat connection with irregular singularity. The solution 
of the flat section equation with such singularity has very interesting Stokes phenomenon, namely,
the asymptotical behavior of the solutions are different in different angular region, and the Stokes
matrix is used to link the solutions in different regions. Those 
Stokes matrices define the generalized monodromy which is essential in describing the moduli space.
The Stokes matrix is a unipotent group which are naturally labeled by a Young Tableaux, and the 
generalized monodromy data can be described by a disc with several marked points labeled by Young Tableaux,
this is exactly the setup we need to apply the construction found in \cite{Xie:2012dw}.

\subsection{Integer pole}
We would like to first discuss the Hitchin equation with irregular singularity
 in some more detail. Let's take a complex structure
on the Riemann surface, the Hitchin equation reads
\begin{eqnarray}
F-\phi\wedge \phi=0, \nonumber\\
D\phi=D*\phi=0,
\end{eqnarray}
where $A$ is the connection and $\phi$ is a one form called Higgs field \cite{Hitchin:1987ab,Kapustin:2006pk}.
By writing ${\cal A}=A+i\phi$, the Hitchin equation implies that the curvature of
${\cal A}$ is flat. In fact, one can introduce a spectral parameter and define 
a family of flat connections.

The monodromy around the singularity can be calculated by solving the
following flat section equation:
\begin{equation}
(\partial_z+{\cal A}_z) \psi=0,
\label{flat}
\end{equation}
with ${\cal A}_z$ the holomorphic part of the complex connection. 
Locally the above equation is just a first order differential equation defined on the  disk.

Let's now turn to irregular singular solution to Hitchin's equation \cite{Witten:2007td}.
The simplest one with gauge group $SU(k)$ is
\begin{eqnarray}
\phi={u_n\over z^n}+.....+{u_2\over z^2}+{u_1\over z}+c.c\nonumber\\
A=\alpha d\theta.
\end{eqnarray}
Here we choose local coordinate $z=re^{i\theta}$, $u_1,...u_n$
are all diagonal matrices after using gauge symmetry, and they are regular semi-simple which means
the eigenvalues are all different.  We ignore the regular terms in the solution but one should keep in mind that
 they are always there. The abelianization of the singular terms  is 
crucial for finding the irregular singular solution. This type of irregular 
singularity is the type I singularity with integer pole we reviewed earlier.

There is an interesting Stokes phenomenon for the differential equation (\ref{flat}) which is important to define the monodromy around the 
irregular singularity.  We will review some aspects for the Stokes phenomenon for the completeness, the interested reader can find
more details in \cite{Witten:2007td,MR919406}. The appearance of 
these  Stokes matrices are coming from the asymptotical behavior of the solutions to the equation $(\partial +{\cal A}_z)\psi=0$.
Assume the gauge group is U(1), then the differential equation  becomes
\begin{equation}
{d\psi\over dz}=-({q_n\over z^n}+{q_{n-1}\over z^{n-1}}...+{q_1\over z}+B(z))\psi,
\end{equation}
here $B(z)$ is a holomorphic function which is regular at $z=0$.  The solution is very simple:
\begin{eqnarray}
\psi=c(z)\exp Q(z),\nonumber\\
Q(z)={q_n\over(n-1) z^{n-1}}+{q_{n-1}\over (n-2)z^{n-2}}...+q_1(-lnz),\
\end{eqnarray}
c(z) is a formal power series which is not convergent around the singularity. This solution can be used to build the solutions for the higher rank group: one 
have a vector of above solution with index $i=1,2,...k$.  However, the entries of solution vector have different asymptotical behaviors 
along different path to the singularity because
\begin{equation}
|{\exp Q^i(z)\over \exp Q^j(z)}|\rightarrow |\exp({q^i_n-q^j_n\over(n-1) z^{n-1}})|,~~~~z\rightarrow 0.
\end{equation}
The asymptotical behavior depends on the sign of $Re({q^i_n-q^j_n\over(n-1) z^{n-1}})$, which are different in different angular regions.  A Stokes ray of
type $(ij)$ is a ray in complex plane where ${q^i_n-q^j_n\over z^{n-1}}$ takes the value in negative imaginary axis:
\begin{equation}
\theta(n-1)=\theta_0+\pi p,
\end{equation}
here $\theta_0=\arg(q^i_n-q^j_n)+{\pi\over2}$ and $n$ is an integer; We also take $0\leq \theta\leq 2\pi$. 
Let's discuss some aspects of Stokes rays' distribution. The angular width between 
two Stokes rays is 
\begin{equation}
\Delta\theta={\pi\over n-1}.
\end{equation}
There are a total of $2(n-1)$ Stokes ray for any pair of solutions. Similarly for any angular regions with
width $\pi\over n-1$, there is a Stokes ray for any pair with the form $(ij)$ and $i> j$, so there are a total of $k(k-1)\over 2$ Stokes rays in this region which are
called Stokes sector. 
\begin{figure}[htbp]
\small
\centering
\includegraphics[width=10cm]{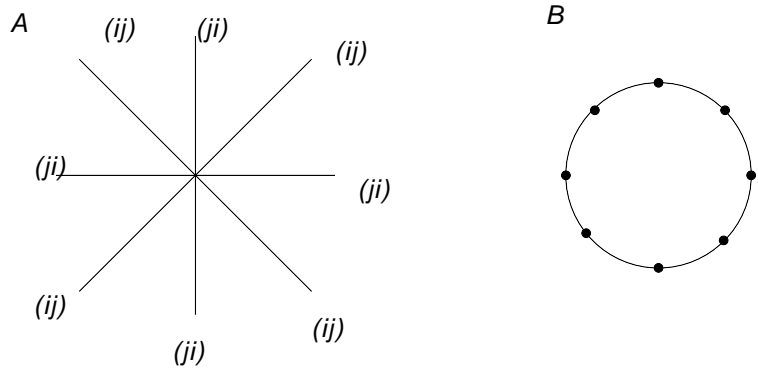}
\caption{A: The Stokes rays for a pair of solutions to meromorphic differential equation on the disc. B: The disc model for the Stokes data of the irregular singularity. }
\label{Stokes}
\end{figure}

The point is that the solution with given asymptotical behavior in a region is not unique  if there is no Stokes ray. For example, if $|{\exp(Q^i(z))\over \exp(Q_j(z)})|>>0$ in this region,
then the solution $\psi_i^{'}(z)=\psi_i(z)+\lambda \psi_j(z)$ has the same asymptotical behavior as $\psi_i$.  Such freedom is not here if there is a Stokes ray in this 
angular region as the asymptotical behavior changes in going through the Stokes ray. 
Let's take a region with angular width $\pi/(n-1)$ whose boundary is not a Stokes ray.  By rotating this region by integer value of $\pi\over n-1$, we get a cover of the 
disk.  One can enlarge each sector a little bit such that there are no Stokes ray in the overlapping region. 
There will be a Stokes ray for any given pair of entries in the solution vector in each sector and the whole solution is uniquely fixed
with given asymptotical behaviors in that region.  On the overlapping region,the two sets of solutions 
are related by an upper triangular matrix with unit diagonal entry.  This matrix is called the Stokes matrix and has the form (by proper renaming the indexes)
\begin{equation}
M=1+\sum_{i<j}c_{ij}E_{ij},
\end{equation}
where $c_{ij}$ is a complex number and $E_{ij}$ is the matrix where the only nonzero entry is $E_{ij}=1$. Explicitly, the Stokes matrix has the following form.

 \[ M=\left( \begin{array}{cccc}
1 & * & * &*\\
0 & 1 & * &*\\
0 & 0 & 1&*\\
 0 & 0 & 0&1\\
\end{array} \right)\]

Each Stokes matrix has a total number of parameters $k(k-1)\over 2$ which equals to the total number of Stokes rays.
The generalized monodromy \footnote{The full monodromy also has a formal monodromy part coming from the regular singular term, and 
the Hitchin's moduli space with irregular singularity is defined by fixing those parameters determining the formal monodromy.}
 is calculated by multiplying $2(n-1)$  Stokes matrices together:
\begin{equation} 
M=M_1*M_2\ldots*M_{2(n-1)}.
\end{equation}
The simple graphic description for the generalized monodromy which are very useful for our later purpose is to blow up the 
irregular singularity to a disc and put $2(n-1)$ marked points on the boundary, each marked point is labeled by a full Young Tableaux $(Y=[1,1,...,1])$.
The reason we assign such a Young Tableaux is determined by the form of the Stokes matrix which is a unipotent subgroup of $SL(k,C)$. 

Let's introduce a little bit group theory which will be useful for our later generalization.  For a diagonal matrix $\alpha$, one could define a parabolic
subgroup ${\cal P}$ which are spanned by elements $\psi$  satisfying
\begin{equation}
[\alpha,\psi]=\lambda \psi,~~~\lambda\geq 0
\end{equation}
The unipotent radical is derived by using the condition $\lambda>0$.  Now let's take $\alpha=diag(y_1,y_2,...,y_k)$ such that $y_1>y_2>...>y_k$,  
then the unipotent radical has the exact same form as the Stokes matrix we just had above. In general, let's use the Young Tableaux to describe 
the degeneracy of the eigenvalues of $\alpha$, i.e.  if the Young Tableaux is $Y=[n_1, n_2,....,n_r]$,  then there are $n_i$ diagonal entries  with same
real value, and we take the same monotonically increasing order for the eigenvalues,  then
the corresponding unipotent radical  is
 \[ S=\left( \begin{array}{ccccc}
I_1 & * & * &*&....\\
0 & I_2 & * &*&....\\
0 & 0 & I_3&*&...\\
 0 & 0 & 0&I_4&...\\
 .&.&.&.&...
\end{array} \right)\]
\label{set}
where $I_j$ is a unit matrix with dimension $n_j\times n_j$.

Now let's relax the condition $u_n$ is regular semi-simple and the irregular singularity is classified by a sequence of Young Tableaux $Y_n\subseteq Y_{n-1} \subseteq.....\subseteq Y_1$. 
This is the type III irregular singularity according to classification. Let's study the solution to the meromorphic differential equation with such an irregular singularity. 
If two solutions have the same order of $z$ dependence and the difference of them is
\begin{equation}
\psi_i-\psi_j={a_{i,j}\over z^{l}}+....
\end{equation}
Then the level $l$ Stokes ray \cite{boalch-2011nn} is defined as the ray in complex plane such that ${a_{i,j}\over z^{l}}$ takes imaginary values.  It is not hard to see that there are a total of $2(l-1)$ Stokes 
ray for each level $l$ pair.  To get the Stokes matrix and the full monodromy, we study the Stokes sector and the corresponding unipotent matrix in linking the solutions of 
two adjacent sectors. 

\begin{figure}[htbp]
\small
\centering
\includegraphics[width=10cm]{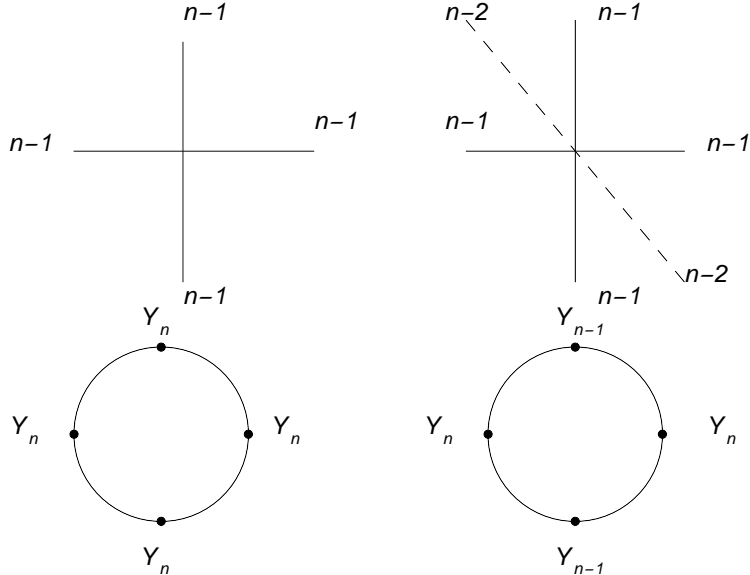}
\caption{A: Stokes data for level $(n-1)$ Stokes rays. B: Level $(n-2)$ Stokes data.}
\label{Stokes2}
\end{figure}

Let's first start with the level $n-1$ Stokes ray, there are $2(n-1)$ Stokes sector following  the above analysis,  with the only exception that the unipotent 
matrix has the generalized form $S$, and the Young Tableaux of this Stokes matrix is $Y_n$, the disc model for level $l$ is shown in figure. \ref{Stokes2}

If $Y_{n-1}\neq Y_{n}$, then some of the columns of $Y_n$ is further partitioned, and we have level $(n-2)$ Stokes ray. However, the total number of Stokes sectors 
are only $2(n-2)$ for this level.  Now there are non-zero entries in the diagonal matrices $I_{n_1}$ and the new Stokes matrix is determined by $Y_{n-1}$. 
However, there are only  $2(n-2)$ matrices determined by 
Young tableaux $Y_{n-1}$,  we are left with two Stokes matrices determined by $Y_n$. One can similarly label the generalized monodromy using the disc model with marked
points on the boundary: there are a total of $2(n-1)$ marked points among which $2(n-2)$ have the label $Y_{n-1}$ and 2 are labeled by $Y_n$.  The cyclic order is 
that there are $(n-1)$ $Y_{n-1}$ marked points followed by one $Y_n$ marked points, and another $(n-1)$ $Y_{n-1}$ followed by the last $Y_n$ label.

\begin{figure}[htbp]
\small
\centering
\includegraphics[width=6cm]{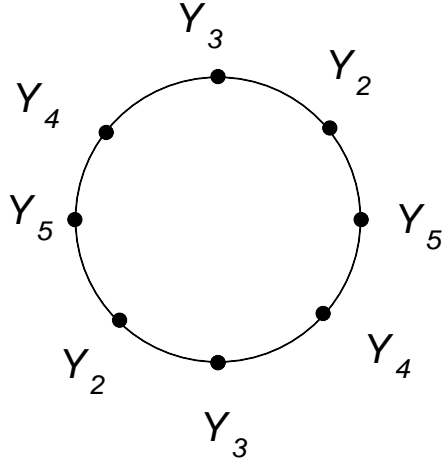}
\caption{The stokes data for type III singularity. }
\label{general}
\end{figure}

One can generalize the above Stokes matrix analysis to other levels, the conclusion is that there are a total of $2(n-1)$ marked points on the boundary: for each Young Tableaux
in the definition of the irregular singularity, there are two marked points sitting on the opposite sides of the line crossing the center of the circle. Moreover,
the cyclic order of the Young Tableaux is exactly $Y_n, Y_{n-1},...Y_2$ on half circle, see figure. \ref{general}.

A special note is that the dimension of the generalized monodromy is equal to the number of Stokes rays, which is equal to the total dimensions of all the Young
Tableaux on the disc boundary if we use the dimension of the corresponding nilpotent orbit labeled by $Y$:
\begin{equation}
dim(Y)={1\over 2}(k^2-\sum n_i^2).
\end{equation}
For the full puncture, we have $dim(Y)={1\over 2}(k^2-k)$ which is exactly the number of Stokes rays in a Stokes sector.

\subsection{Type I irregular singularity}
Let's recall the form of this type of irregular singularity which will appear in the definition of the meromorphic connection:
\begin{equation}
\Phi={1\over z^{n+j/k+2}}diag(1,\omega,\omega^2,...\omega^{k-1})+...
\end{equation}
As we noted above, the key is to study the number of Stokes ray and its distribution. 
The Stokes phenomenon for this type of irregular singularity can be understood by going to 
the covering space: let's define $z=t^k$, then the irregular singularity has the form
\begin{equation}
\Phi={1\over t^{(n+1)k+j+1}}diag(1,\omega,\omega^2,...\omega^{k-1})+..
\end{equation}
We ignore an overall real factor which is not important here.
This irregular singularity is the familiar type where the leading order is regular semi-simple. 
The analysis of the Stokes phenomenon and Stokes matrices is now straightforward, however, one should 
note that not all the Stokes rays on the covering space are the actual Stokes ray on the original plane.
Let's consider a pair of solutions corresponding to two eigenvalues of the irregular connection and 
assume the difference of them has the angular coordinate $\theta_{0}$, then the condition for 
the Stokes ray on the original space is
\begin{equation}
\theta_1(n+j/k+1)=\theta_0+\pi p
\end{equation}
with $p$ an integer and $0\leq\theta_1\leq 2\pi$. On the other hand, the condition for the Stokes ray on the covering space is 
\begin{equation}
\theta_2((n+1)k+j)=\theta_0+\pi q
\end{equation}
with $q$ an integer and $0\leq\theta_2\leq 2\pi k$. So among $k^2$ Stokes rays in the covering space, only 
one is the real Stokes ray. Therefore the  angular width of the Stokes sector is ${\pi k^2\over (n+1)k+j}$. 
So we do not have an integer number of Stokes sector in the covering space.

What we are going to define the Stokes sector is the following: consider first $2(n+1)$ maximal Stokes sector, and then $2j/k$ portion of the 
maximal Stokes sector which we call it $S_0$. Since the Stokes rays are distributed equally in the covering space, we conclude
that the total number of Stokes ray on this sector is
\begin{equation}
N_{S_0}=j(k-1),
\end{equation}
which can also be found by explicit calculations in the original space.
Since the minimal unipotent group has dimension $(k-1)$, and the fractional part of Stokes sector can be replaced by 
$j$ simple Stokes matrices whose Young Tableaux is $[k-1, 1]$.

\begin{figure}[htbp]
\small
\centering
\includegraphics[width=6cm]{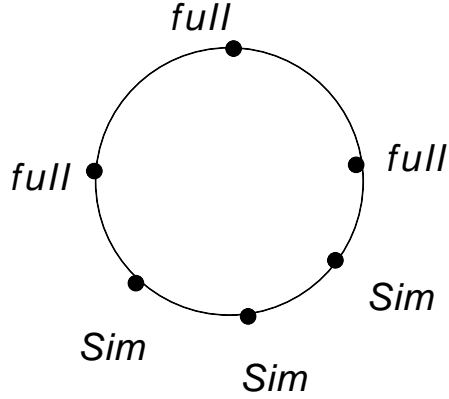}
\caption{The stokes data for type 1 singularity. }
\label{type1}
\end{figure}

The total number of Stokes rays are
\begin{equation}
N_{Stokes}=k(k-1)(n+1)+j(k-1),
\end{equation}
which is the total number of parameters for the generalized monodromy.
The disc model for this type of irregular singularity is the following: one first have $2(n+1)$ 
full punctures and then $j$ simple punctures distributed on the disc in a cyclic order, see figure. \ref{type1}.
When $k$ is even, and $j={k\over 2}$, one could equivalently put $2n+3$ full punctures on 
the boundary.

\subsection{Type II irregular singularity}
The Higgs field of this irregular singularity has the form
\begin{equation}
\Phi={1\over z^{n+j/(k-1)+2}}diag(0,1,\omega,\omega^2,...\omega^{k-2})
\end{equation}
with $\omega=\exp({2\pi i\over k-1})$. The Stokes ray of a given pair of solutions is defined by the condition
\begin{equation}
\theta(n+1+j/(k-1))=\theta_0+\pi p.
\end{equation}
The angle difference between two Stokes rays is
\begin{equation}
\Delta\theta={\pi\over n+1+j/(k-1)}.
\end{equation}
The total number of Stokes rays are
\begin{equation}
N_{Stokes}=(n+1)k(k-1)+kj.
\end{equation}

\begin{figure}[htbp]
\small
\centering
\includegraphics[width=6cm]{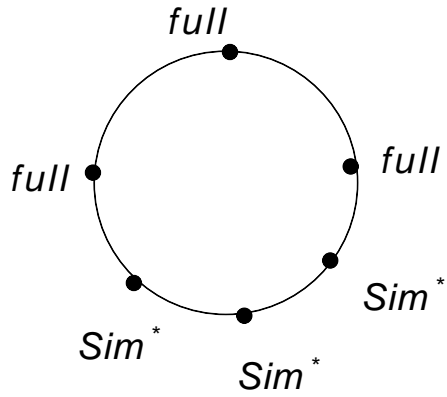}
\caption{The disc model for type II singularity, here we need to include the boundary node from the simple puncture. }
\label{type2}
\end{figure} 

The difference with type I singularity is that there is an extra $jk$ instead of $j(k-1)$ number of Stokes rays.  There is no simple 
representation using the Young Tableaux, but we have the following proposal:
The disc model for 
such singularity is the same as the previous one, but we want to include the boundary nodes for 
the simple puncture, see figure. \ref{type2}.  This proposal will be confirmed in our later study, it would be interesting to prove 
it though. If $k$ is odd, and $j={k-1\over 2}$, one can put $(2n+3)$ full punctures
on the boundary to represent the monodromy data  instead of gauging the boundary node.

\newpage

\section{Cluster coordinates for Argyres-Douglas theory}
The generalized monodromy data for the irregular singularity is denoted by a sequence of Young Tableaux distributed on 
the boundary of the disc. The corresponding cluster coordinates are going to be constructed 
using the idea developed in \cite{Xie:2012dw}. In that paper, given a cyclic sequence of the Young Tableaux on the boundary ,
we can construct the network (equivalently quiver) from the ideal triangulation. We are going to show that the 
quiver actually has the same dimension as the rank of the charge lattice of the field theory, therefore, these coordinates 
are actually describing the moduli space of the framed local system in the presence of the irregular singularity. And we 
conjecture that the quiver is the corresponding BPS quiver.

This idea actually has firm grounding in the case when the irregular singularity has integer pole with leading order
regular semi-simple. It is shown that the partial monodromy for a path around one vertex of the triangle \cite{fock-2003} is actually 
upper-triangular which has exactly the same form as the Stokes matrix. It is interesting to verify that this fact is also 
true for the general punctures. 

Let's discuss more about the ideal triangulation of Riemann surface with boundaries since the 
irregular singularity is replaced by a boundary with labeled marked points. Let us begin with a Riemann surface with boundaries,
and specify a finite set of points $M_{\rm boundary}$,
called boundary marked points,
on the boundary circles of $\Sigma$.
Each connected component of $\partial \Sigma$ has at least one
boundary marked point;  The bulk puncture is not blown up and remained as a point in the interior of the Riemann surface.
The defining data of our theory is a triple
$(\Sigma, M_{\rm boundary},p)$, 
For notational convenience we
sometimes denote this triple simply by $\Sigma$. 
In other words, $\Sigma$ is defined by following data:

a. the genus $g$ of the Riemann surface;

b: the number of bulk punctures $p$.

d. the number $b$ of boundary components;

d. the number of marked points $h_i$ on each boundary.

Each puncture represents the regular singularity while the boundary with marked points 
means an irregular singularity, all the marked points have a Young Tableaux label. The punctures and the marked
points are all called marked points for simplicity in the following, and one should be careful about whether it is in the bulk or one the boundary.
One can define a combinatorial object called ideal triangulation on above Riemann surface.
 An ideal triangulation is defined using arcs \cite{fomin-2008-201}.  A simple arc $\gamma$ in $\Sigma$ is a curve such that 

1. the endpoints of $\gamma$ are marked points; 

2. $\gamma$ does not intersect itself, except at the endpoints;

3. $\gamma$ is disjoint from the marked points and the boundary.

We also require the arc $\gamma$ is not contractible into the marked points or onto the boundary. Each arc is considered up to
isotopy. Two arcs are called compatible if they do not intersect in the interior of $\Sigma$.  A maximal collection of distinct pairwise arcs 
is called an {\it ideal triangulation}. An edge is called external if it is isotopic to a segment of the boundary, otherwise
it is called internal.

It is not hard to get the following formula for the number of internal edges:
\begin{equation}
6 g+3b+3p+\#\left| M_{\rm boundary }\right|-6 \ ,
\end{equation}
where as defined previously $g$ ($b$) is the genus (the number of boundary components) of
$\Sigma$, respectively.
The number of internal edges is $6g+3b+3p-6$, and there are 
a total of $\#\left|M_{\rm boundary}\right|$ external edges.

Once the ideal triangulation is given, one need to further tesselate the triangles to construct a dot diagram 
according to the marked points type, and then find the network and quiver. In this section, 
we will give a complete analysis of all the Argyres-Douglas constructed in \cite{Xie:2012hs}, which involves at most
one irregular singularity on the Riemann sphere, so we have only one boundary component.
In the next section, we would consider general boarded Riemann surface which represents
the asymptotical free theory.

\subsection{Type I AD theory}

\subsubsection{$(A_1,A_{N-1})$ theory}
Let's first study the construction of quiver for $(A_1, A_{N-1})$ theory using network (which is 
equivalent to the triangulation).  The quiver for this theory is well known and our construction gives another combinatorial object: network which
turns out to be very useful. The six dimensional construction involves a type I irregular singularity of 
$A_1$ group.  The order of pole is $\lambda={N\over 2}+2$ so the number of marked points on the 
boundary is $n_{marked}=N+2$ according to our previous analysis on Stokes data.  
The bordered Riemann surface is a disc with $(N+2)$ marked points. There is only one type of puncture for
$A_1$ theory, and all the marked points have the same type.

\begin{figure}[htbp]
\small
\centering
\includegraphics[width=10cm]{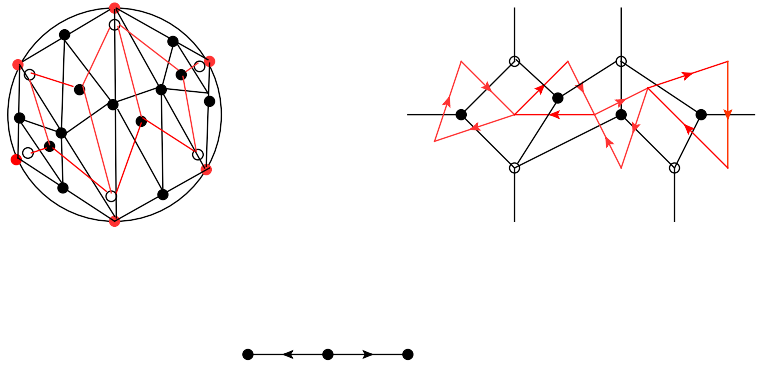}
\caption{Left: the triangulation and the network for $(A_1, A_3)$ theory. Right: The planar network and the quiver. Bottom: the quiver from the network and we keep only the nodes from the closed surface.}
\label{A1AN}
\end{figure}

\textbf{Example 1}: The ideal triangulation of the disc with six marked points is given in figure. \ref{A1AN}, which 
describes $(A_1, A_3)$ theory. The network  can be 
easily constructed and the quiver 
is also given in figure. \ref{A1AN} which is indeed of $A_{3}$ shape \footnote{The orientation of the quiver arrows of this quiver is not important, since quivers with different orientations are
related by quiver mutation.}. 
 Notice that the open surface 
on the boundary is not included as the quiver node, only the closed surface is used.  Now 
we can do square moves and produce equally good coordinates, such square move 
is just a graphical representation of quiver mutation.

Let's give some general topological descriptions of the network.  We add one external edge to the white vertices on the vertex of the triangulation and 
 delete vertices with only two edges (which will not change the quiver). So there are a total of $N+2$ external edges which is equal to the number of punctures. For
 a general network, define $N$ as the total number of the punctures, then
one can find  an integer $K$ from the following formula :
\begin{equation}
K-(N-K)=\sum {\rm col}(v)({\rm deg}(v)-2) \ ,
\label{kdef}
\end{equation}
where ${\rm col}(v)=1$ for black vertices and ${\rm col}(v)=-1$ for white vertices. One could verify that 
$k$ is not changed under quiver mutation and therefore this is a quiver mutation invariant. 
Each such network describes a positive cell in Grassmannia  $G(N,K)$. 
It is easy to verify for $(A_1, A_{N-1})$ theory, $K$ is always equal to $2$, and the network 
describes maximal positive cell of  $G(N+2,2)$.  Moreover, one could define the permutation using the zig-zag path which 
is also invariant under the square move.

\subsubsection{$(A_{k-1},A_{nk-1})$ theory}
The next simplest case is the $(A_{k-1}, A_{nk-1})$ theory whose network can be found easily.  The irregular singularity for this theory is 
\begin{equation}
\Phi={A_r\over z^r}+{A_{r-1}\over z^{r-1}}+\ldots+{A_1\over z},
\end{equation}
with $r=n+2$, and $A_r, A_{r-1},...,A_1$ all have distinct eigenvalues.
According to our previous analysis of the Stokes phenomenon, there are a total of 
$2(n+1)$ Stokes matrices arranged on the disk around the singularity. Each Stokes matrix is labeled by 
a  full Young tableaux with partition $[1,1,\ldots,1]$. 

Geometrically, the Riemann sphere with irregular singularity is replaced by a disc with $(2n+2)$ marked points 
on the boundary. The triangulation of the disc has $(2n+2)$ external edges and $(2n-1)$ internal edges.
The network is constructed from a triangulation and dot diagram on each triangle. The quiver is actually easy to
find just from the dot diagram.

The total number of quiver nodes can be counted as following: the number is equal to the  Coulomb 
branch dimension of the ${\mathcal N}=2$ theory defined by putting the same number of full punctures on the sphere, which
is equal to the sum of the dimension of all the Young tableaux minus the dimension of the gauge group
\begin{equation}
N={1\over2}2(n+1)(k^2-k)-(k^2-1)=(k-1)(nk-1).
\end{equation}

Amusingly, as shown in formula $[6.21]$ of \cite{Xie:2012hs}, the rank of the charge lattice is
\begin{equation}
R=(k-1)(nk-1),
\end{equation}
which is equal to the rank of the quiver we just constructed. So we conjecture that the quiver from the network
is the BPS quiver for the underlying field theory.

\begin{figure}[htbp]
\small
\centering
\includegraphics[width=10cm]{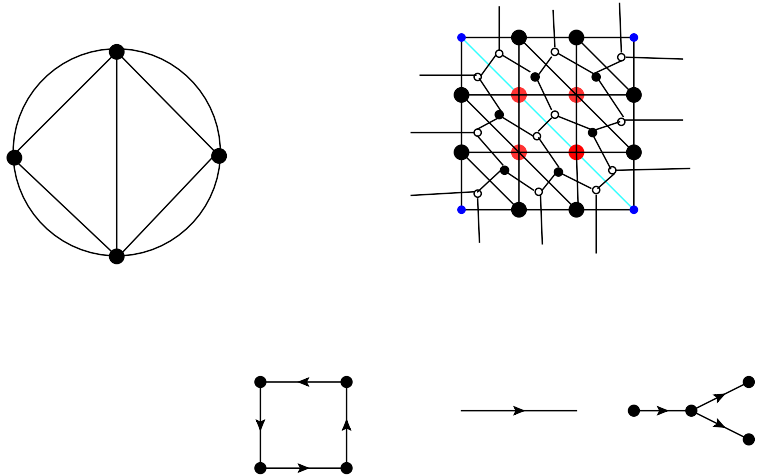}
\caption{The network and the quiver for $(A_2,A_2)$ theory. The quiver is mutation equivalent to $D_4$ Dynkin 
diagram. }
\label{A2A2}
\end{figure}

\begin{figure}[htbp]
\small
\centering
\includegraphics[width=10cm]{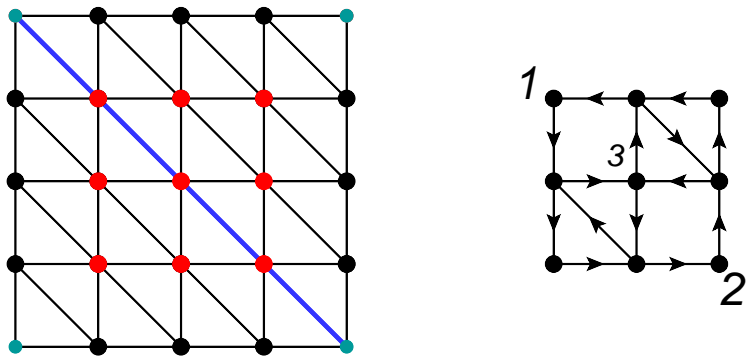}
\caption{The tessellation for $(A_3,A_3)$ theory and the quiver  is mutation equivalent to $A_3$ times $A_3$ Dynkin diagram.
This can be shown by first mutating nodes $1$ and $2$, then on node $3$.}
\label{A3A3}
\end{figure}

\textbf{Example 2}:   Let's consider $(A_2, A_2)$ theory, there are a total of 4 marked points on 
the boundary of the disc, and each marked point is labeled by a full puncture of $A_2$ group.  The 
triangulation is very simple and the dot diagram in each triangle is easily found.
The quiver has the shape of  $A_2\times A_2$, and confirm the result presented  in \cite{Cecotti:2010fi}.
The resulting quiver diagram is mutation equivalent to the $D_4$ Dynkin diagram, which is a further justification 
that the $(A_2, A_2)$ theory is isomorphic to $(A_1, D_4)$ theory as shown in figure. \ref{A2A2} \footnote{We do not draw the network explicitly for most of the examples below.}. This isomorphism can also be checked from 
various other tests like the scaling spectrum, the central charges, three dimensional mirror, etc. The quiver for $(A_3, A_3)$ theory 
is shown in figure. \ref{A3A3}. If we first do quiver mutation on node $1$ and $2$, then do the quiver mutation on $3$, 
the final quiver is of the shape of a product of $A_3$ and $A_3$ Dynkin diagram. The interested reader can check the quiver is mutation 
equivalent to a product of two Dynkin diagrams for the general cases.

After finding the network from the triangulation, one could find other quivers by doing square moves which 
are just the quiver mutation acting on the quivers. In particular, the network for different triangulations are related by 
a sequence of square moves. 

Let's now find the invariant information associated with the network which is not changed under square move. One can find the permutations using zig-zag path which 
is invariant under the square move, and we do not have anything more to say about this. It is interesting to find the underlying
Grassmannia though.  The computation can be done by explicitly counting the degree of the various vertices.

First, there are a total of  $(2n+2)$ external edges and $(2n-1)$ internal edges in the triangulation; There are a total of $2n$ triangles.
The $N$ of the network is easy to count: there are a total $(k-1)(2n+2)$ external edges in the network, so 
\begin{equation}
N=(k-1)(2n+2).
\end{equation}

Let's now count the degree of various vertices.
There are ${k(k-1)\over 2}$ trivalent black vertices and ${(k-3)(k-2)\over 2}$ trivalent white vertices in each triangle.
The white vertices on the boundary are a little bit complicated. The degree of white vertices on the internal edge
is four, and the total number of such vertices are $(2n-1)(k-2)$; the white vertices on the external edge which is not living at the marked points has degree 3,  and 
the total numbers are $(2n+2)(k-2)$.   The white vertices at the marked points on the boundary can be counted using the special zig-zag 
triangulation, and there are two degree two vertices, two degree three vertices, and $(2n-2)$ degree four vertices, so the total
contribution of these white vertices are $2(2n-1)$. Combining the above 
analysis, we have the following formula
\begin{equation}
K-((k-1)(2n+2)-K)=2n[{k(k-1)\over 2}-{(k-3)(k-2)\over 2}]-(2n+2)(k-2)-2(2n-1)(k-2)-2(2n-1),
\end{equation}
we find $K=k$. So the corresponding Grassmannia is $G(2(n+1)(k-1),k)$.

\newpage

\subsubsection{Generic case}
Let's now consider general $(A_k, A_{nk-1+j})$ theory with $ 0<j<k$.  The Stokes matrices analysis suggests that there are 
$j$ more marked points which are labeled by simple Young Tableaux.  The number of marked points and their labels are
\begin{equation}
full: 2(n+1);~~~~~simple: j
\end{equation}
The full punctures are grouped together.

\textbf{Example 2} : Consider $(A_{k-1}, A_1)$ theory,  here we have $n=0$ and $j=2$ and
there are two full punctures and two simple punctures. The quiver can be found using the tessellation of the quadrilateral bounded by two full
punctures and two simple punctures.  The quiver has indeed the shape  of $A_{k-1}$ Dynkin diagram.
\begin{figure}[htbp]
\small
\centering
\includegraphics[width=10cm]{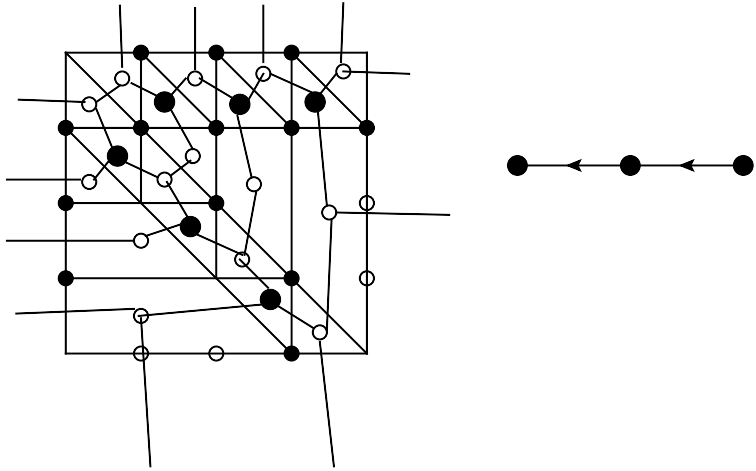}
\caption{The network and the quiver for $(A_3,A_1)$ theory. The  quiver has the shape of $A_3$ Dynkin diagram.}
\label{AkA1}
\end{figure}

\begin{figure}[htbp]
\small
\centering
\includegraphics[width=8cm]{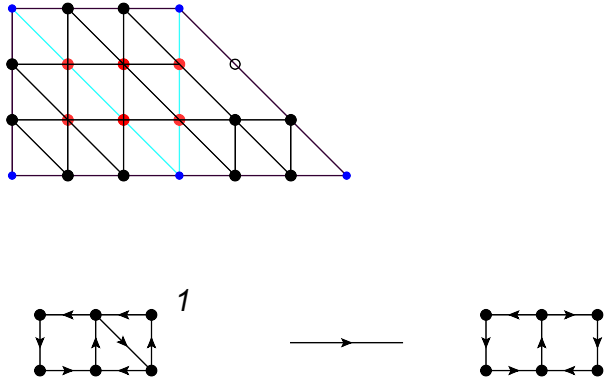}
\caption{The tessellation and quiver for $(A_2, A_3)$ theory, we have ignored the network and write the quiver directly. It is mutation equivalent to $(A_2, A_3)$ quiver
if we mutate the quiver node 1. One can also show that this quiver is mutation equivalent to $E_6$ Dynkin diagram.}
\label{E6}
\end{figure}

\textbf{Example 3}: Consider $(A_2, A_3)$ theory which is also isomorphic to $(A_1, E_6)$ theory. This theory is engineered using
$A_2$ theory with $n=1$ and $j=1$ . There are four full punctures and one simple puncture according to our discussion of the Stokes phenomenon.
The network and quiver are shown in figure. \ref{E6}. The quiver from the network is not of the $E_6$ shape, but 
one can use the Java program \cite{Keller:2012}  to show that it is mutation equivalent to $E_6$ 
Dynkin diagram \footnote{First draw the quiver, and select tools and mutation class, you can find the simplest quiver in the mutation class.} .

\begin{figure}[htbp]
\small
\centering
\includegraphics[width=8cm]{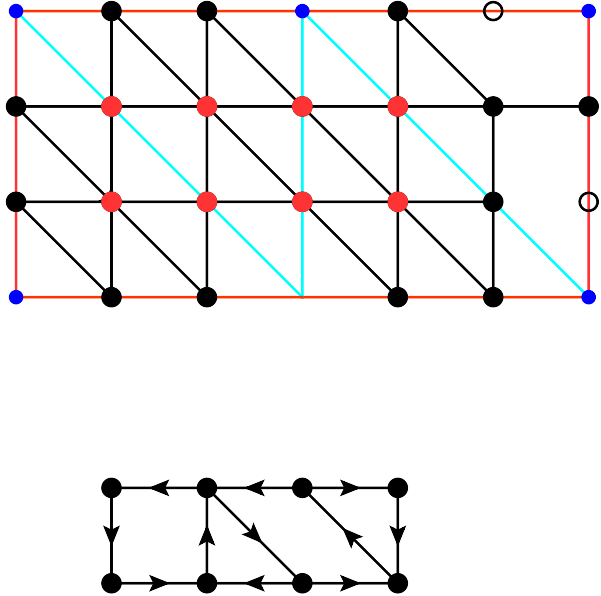}
\caption{The tessellation and quiver for $(A_2, A_4)$ theory; This quiver is not obviously mutation equivalent to quiver of $A_2$ Dynkin diagram times $A_4$ Dynkin diagram though.
However, this quiver is mutation equivalent to $E_8$ Dynkin diagram by using the Java program, and $E_8$ quiver is also mutation equivalent to $(A_2,A_4)$ theory. }
\label{E8}
\end{figure}

\textbf{Example 4}:  This example deals with $(A_2, A_4)$ theory which is expected to be isomorphic to $(A_1, E_8)$ theory. There are 
four full punctures and two simple punctures on the boundary of the disc. The dot diagram and the quiver is shown in figure. \ref{E8}. Using the Java program, it can be shown that the 
quiver is mutation equivalent to $E_8$ Dynkin diagram.

The network constructed is always a minimal network and we would also like to get some invariant information.  The total external edges 
are $(2n+2)(k-1)+j$.  We also find that $K=k$ for many examples, so the corresponding Grassmannia is $G((2n+2)(k-1)+j, k)$.  This seems to be 
the general result: the network  from  $A_{k-1}$ theory on a disc always describe the Grassmannia $G(.,k)$, it would be interesting to prove this fact.

\begin{figure}[htbp]
\small
\centering
\includegraphics[width=12cm]{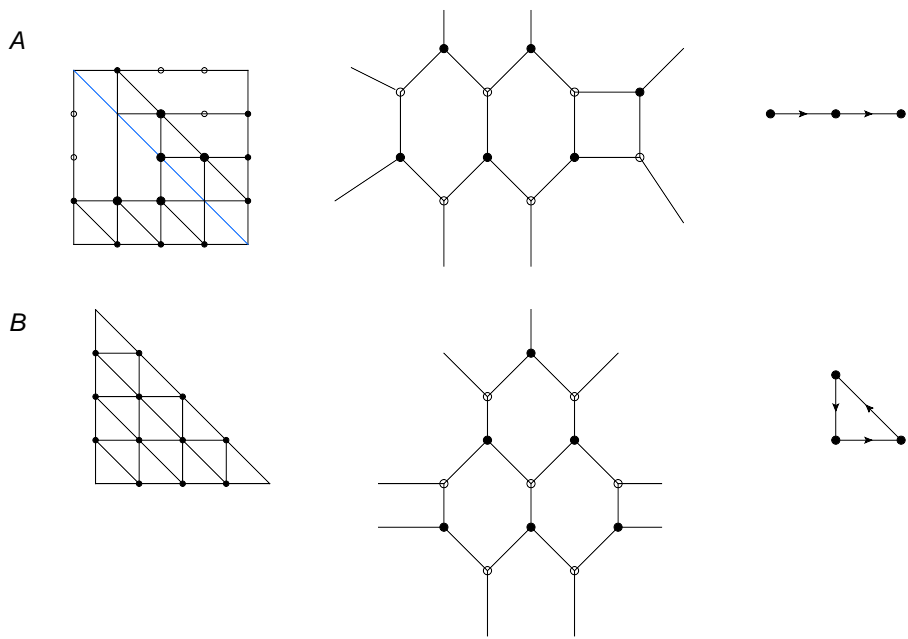}
\caption{A: The network and quiver for $(A_3, A_1)$ theory by using the simple punctures. B: The quiver derived using
the full puncture. These two quivers are in the same mutation class. }
\label{Equivalence}
\end{figure}

If $k$ is even and $j=k/2$, then there is another representation where we replace the $j$ simple punctures with a full puncture. 
These two choice lead to two different quivers which are in the same mutation class.  Let's also consider $(A_3, A_1)$ theory, the two representations 
give the quivers in the same mutation class, see figure. \ref{Equivalence}. However, the networks for two representations are not in the same class, i.e. they
are not related by square move . 

Given the cyclic choices of the punctures, there are actually two ways of decorating the edges: clockwise or anti-clockwise \cite{Xie:2012dw}. However,
in our current situation, they yield the same quiver mutation class. The idea is the following: consider the simple puncture and 
the triangulation where there are no internal edges ending on this puncture (this does not lose any generality since the quivers for
different triangulations are related by the quiver mutations), then the effect of the special puncture is to cut a $(k-1)\times(k-1)$ 
triangle around this puncture, and this fact does not depend on how you choose to decorate the edges! 

\newpage
\newpage
\subsection{Type II AD theory}
The type II AD theory is defined on a Riemann sphere with the following irregular singularity: 
\begin{equation}
\Phi={A_r\over z^r}+\ldots,
\end{equation}
where $ r=n+{j\over k-1}+2 $ and $A_r$ has the form
\begin{equation}
{A_r}=diag(0, 1, \omega,...\omega^{k-2}),
\end{equation}
with $\omega=exp({2\pi i\over k-1})$.  The total number of parameters from the Stokes matrix (equivalently total number of Stokes ray) is 
\begin{equation}
N_{Stokes}=(n+1)k(k-1)+kj.
\end{equation}
The total number of quiver nodes is the above number minus the dimension of the gauge group. 
Not surprisingly, the rank of the charge lattice of the corresponding AD theory has exactly same 
number by just counting the Coulomb branch and mass parameters. So we conjecture the quiver is the BPS quiver for the corresponding AD theory. 
\begin{figure}[htbp]
\small
\centering
\includegraphics[width=8cm]{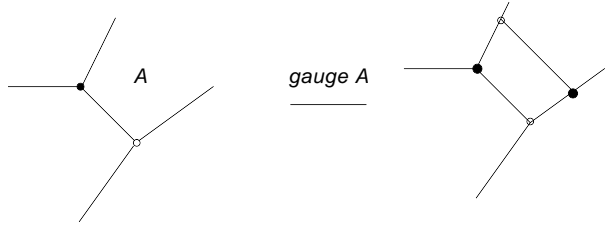}
\caption{The open surface of the network is gauged by adding an extra edge.  }
\label{gauge}
\end{figure} 

The network is constructed from the triangulation of the disc with several marked points. There are $2(n+1)$ full punctures and $j$ simple puncture.
The boundary node of the simple puncture is going to be gauged. At the level of the network, we add an extra edge to close the open surface corresponding
to the node on the simple puncture edge as shown in figure. \ref{gauge} \footnote{This is like adding a  BCFW bridge in the study of the scattering amplitude.}. 
One interesting fact is that the network produced after this operation is still the minimal network and the Grassmannia cell is still of the form 
$G(., k)$.

\begin{figure}[htbp]
\small
\centering
\includegraphics[width=8cm]{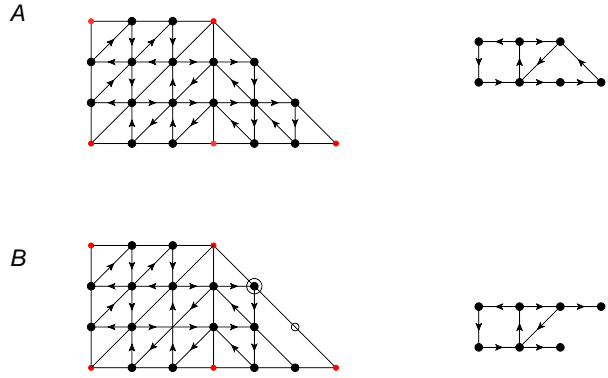}
\caption{A: The triangulation of the disc of the $E_7$ theory using the full puncture representation. B: Another derivation using the gauging idea.  
One can show that these two quivers  are mutation equivalent to each other using the Java program. }
\label{E7}
\end{figure}

\textbf{Example 1}: $E_7$ theory: The irregular singularity for this AD theory has $k=3$ and $n=1, j=1$, there are $5$ full punctures on the disc,
the triangulation and the corresponding quiver is shown in figure. \ref{E7}. The quiver from our construction is not as the form of $E_7$ Dynkin diagram,
but after several quiver mutations, one can find that the quiver is indeed of the E7 shape. There are other realizations in which we have four full punctures,
and one simple puncture which are gauged. The dot diagram and the corresponding quiver is shown in figure. \ref{E7}. The quiver is also 
mutation equivalent to the $E_7$ quiver by using the Java program.

\textbf{Example 2}: The irregular singularity of this theory has $n=0, j=2$ which is equivalent to $(A_1, D_N)$ theory as shown in \cite{Xie:2012hs}.
There are two full punctures and two simple punctures which are going to be gauged. The resulting quiver is mutation equivalent to $D_N$
Dynkin diagram which further confirms our previous identification.

\begin{figure}[htbp]
\small
\centering
\includegraphics[width=8cm]{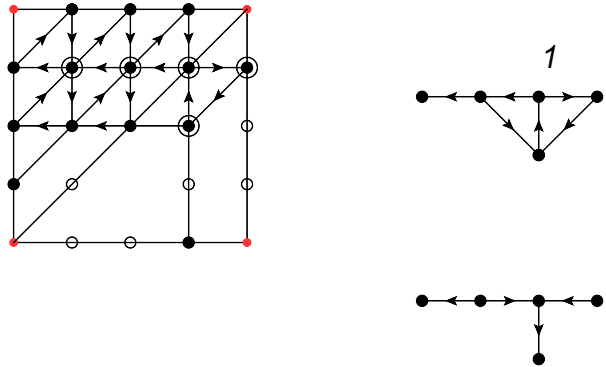}
\caption{This type II network leads to $D_N$ quiver. }
\label{DN}
\end{figure} 

\newpage
\subsection{Type III AD theory}
This theory is specified by a sequence of Young Tableaux $Y_n\subset Y_{n-1}....\subset Y_1$.  The charge lattice of the corresponding 
AD theory has dimension:
\begin{equation}
R=2(\sum_{i=n}^{i=1}dim(Y_i)-dim(G))+r,
\end{equation}
where $r$ is the number of heights of the Young Tableaux $Y_1$.

The total number of Stokes rays are
\begin{equation}
N_{Stokes}=2\sum_{i=n}^{i=2}dim(Y_i)
\end{equation}
where $dim(Y_i)$ is the dimension of the nilpotent orbit labeled by Young Tableaux $Y_i$.  Now the disc model of this irregular singularity
have twice number of Young Tableaux of $Y_2$ to $Y_n$, and the ideal triangulation will give the following number of quiver nodes
\begin{equation}
N_{nodes}=2\sum_{i=n}^{i=2}dim(Y_i)-dim(G).
\end{equation}
We want this number to be the same as the dimension of the charge lattice of the field theory, and $Y_1$ should satisfy the following constraint: 
\begin{equation}
2dim(Y_1)+r=dim(G),
\end{equation}
which is only possible if $Y_1$ is a full Young Tableaux. This is the case we are going to discuss in this section.

\begin{figure}[htbp]
\small
\centering
\includegraphics[width=8cm]{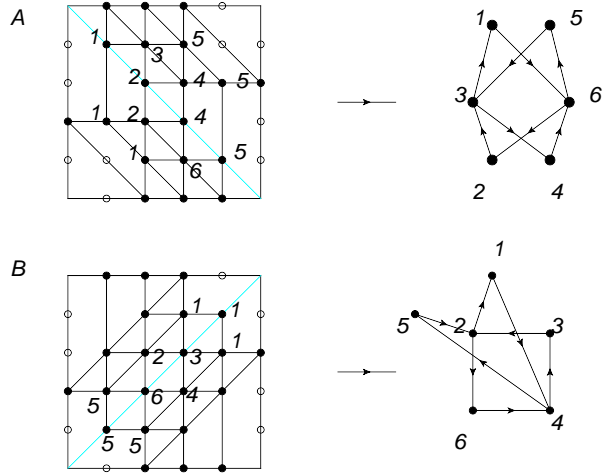}
\caption{The triangulation and  the quiver for a type III singularity with $Y_2=[2,3]$ and $Y_1=[2,1,1,1]$. The quiver is the same as 
the that of the $SU(2)$ with four flavors.}
\label{typeIII}
\end{figure}

\begin{figure}[htbp]
\small
\centering
\includegraphics[width=8cm]{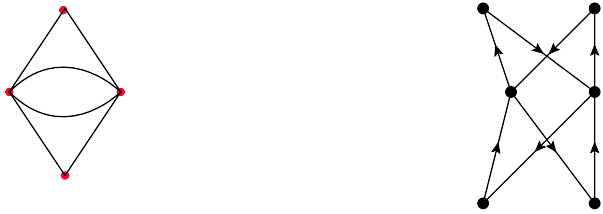}
\caption{The quiver for $SU(2)$ with four flavors, which comes from the triangulation of the sphere with four full punctures of $A_1$ theory.}
\label{SU2}
\end{figure}

\textbf{Example 1}: Let's take $Y_2=[2,3]$ and $Y_1=[2,1,1,1]$. Then the tessellation and the quiver is shown in figure. \ref{typeIII}.
The two networks are not minimal as discussed in \cite{}, this is one of the big difference with the  network discussed in 
Grassamannia issue. It seems that the BPS quiver does not need the minimality condition, in fact, we are just care about the closed surface.
There are two triangulations and the quiver is quiver mutation equivalent, although the underlying network is not square equivalent
to each other, the quivers are mutation equivalent. According to the discussion in \cite{Xie:2012hs}, this theory is isomorphic to 
the $SU(2)$ with four flavors whose BPS quiver is read from a sphere with four simple punctures of $A_1$ theory.
The BPS quiver from that representation is shown in figure. \ref{SU2}, we can see that the rank $5$ representation is indeed giving the 
same BPS quiver.

There is one question I would like to comment. According to the analysis presented in \cite{Xie:2012dw}, the quiver from different
 triangulations are not related by the square move if the maximal height of  the first Young Tableaux is bigger than one, 
 and the minimal height of the next Young Tableaux in the cyclic order is also bigger than one.  In our current example, the quivers from 
 two triangulations are mutation equivalent if we only consider the internal nodes. We suspect this is true for all the triangulations and 
 punctures. 

\subsection{Type IV AD theory}
One could add another regular puncture to above irregular puncture, then the corresponding Riemann surface is a disc with marked points and one 
bulk puncture. The triangulation of this Riemann surface can be easily found, the decoration of the external edge is the same as the previous case, and
One of edge coming into the bulk puncture should be decorated using the Young Tableaux.  The total number of quiver nodes from the Young Tableaux
is the same as the charge lattice of the field theory. Let's work out some examples in detail. 

\begin{figure}[htbp]
\small
\centering
\includegraphics[width=8cm]{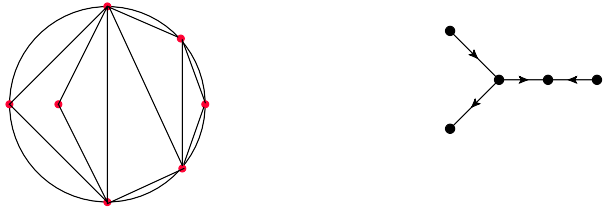}
\caption{The triangulation and quiver for $(A_1, D_{N+2})$ theory.}
\label{DN1}
\end{figure}

\textbf{Example 1}: The $(A_1, D_{N+2})$ theory is engineered from six dimensional $A_1$ theory on a sphere with one irregular singularity and 
one regular singularity.   The bordered Riemann surface is just a disc with several puncture and one bulk  puncture. One triangulation and the 
quiver is shown in figure. \ref{DN1} which is exactly of the D type Dynkin diagram shape and justifies the name of the theory. One could also 
write a planar network and the network is not minimal precisely because of the presence of the bulk puncture. 
The simple higher rank generalization is to replace each puncture with the full puncture. The triangulation is the same and 
each triangle has internal structures, the full quiver is straightforward to find. More generally, one could use any of the irregular singularity 
studied in previous subsection and find the corresponding BPS quiver pretty easily.

\begin{figure}[htbp]
\small
\centering
\includegraphics[width=8cm]{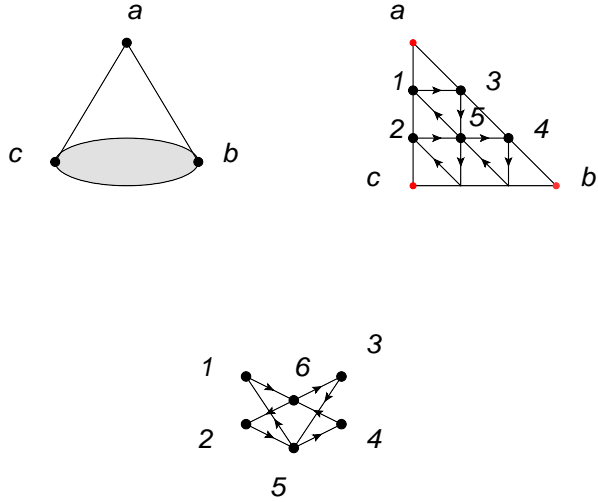}
\caption{ The triangulation of a disc with two full marked points and one full bulk puncture of $SU(3)$, the quiver is in the same mutation class of the $SU(2)$ with four flavors.}
\label{R3}
\end{figure}

\textbf{Example 2}: This example is derived from using six dimensional $A_2$ theory, one has two regular puncture on the boundary of the disc, and 
another full bulk puncture, the triangulation and the corresponding tessellation is given in figure.\ref{R3}. We find the quiver and it is the same as the $SU(2)$
with four flavors, which confirms the observation made in \cite{Xie:2012hs}.

\begin{figure}[htbp]
\small
\centering
\includegraphics[width=10cm]{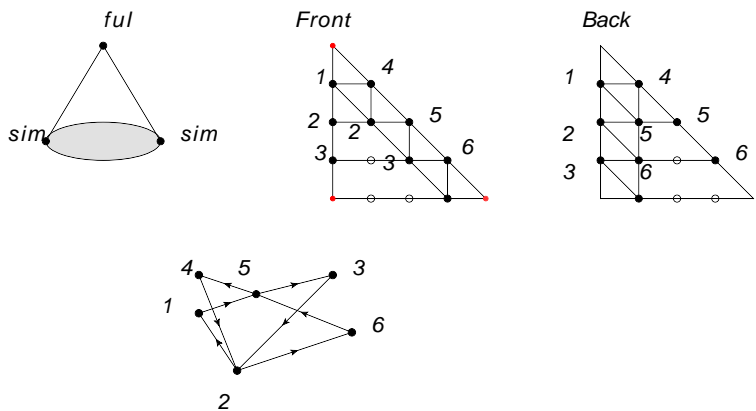}
\caption{ The quiver for a Type IV AD theory using $A_4$ theory which is actually isomorphic to $SU(2)$ with four flavors. }
\label{rank4}
\end{figure}

\textbf{Example 3}: Now let's consider $A_3$ theory and also two simple punctures on the boundary of the disc, and the bulk puncture is full.
The triangulation and the quiver is given in figure. \ref{rank4}, it is remarkable that it is also the same as the $SU(2)$ with four flavors and this 
is a further justification of our claim made in \cite{Xie:2012hs}.

\newpage
\section{Cluster coordinates for asymptotical free theory}

\subsection{Linear quiver  with Lagrangian descriptions}

\subsubsection{One gauge group}
Let's first consider $SU(N)$ SQCD with various number of flavors.  The six dimensional construction involves two irregular singularities which 
actually represent the number of flavors attached on the gauge group. The  irregular singularity for $n_f$ fundamental is \cite{Nanopoulos:2010zb}:
\begin{equation}
\Phi={1\over z^{1+{1\over k-n_f}}}diag(0,..0,1,\omega,...\omega^{k-n_f-1})+{1\over z}diag(m_1,....m_{n_f}, M,M, \ldots, M).
\end{equation}
Here $\omega=exp({2\pi i\over k-n_f})$.
When $n_f=k-1$, one have a type III irregular singularity with Young Tableaux $Y_2\subset Y_1$, here $Y_2$ is a simple and $Y_1$ is a full Young Tableaux.

\begin{figure}[htbp]
\small
\centering
\includegraphics[width=6cm]{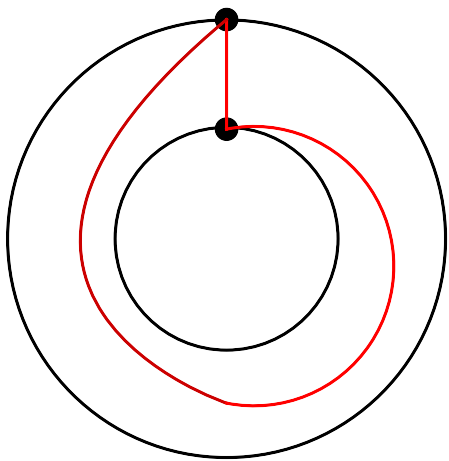}
\caption{ Triangulation of the annulus relevant for pure SYM theory.}
\label{annulus}
\end{figure}

The pure SYM case is rather simple, since there are two irregular singularities with the same type as the type I irregular singularity with $n=-1$ and $j=1$.
The Riemann surface is a annulus with one marked point on each boundary which is of the simple type.
 The triangulation of the annulus is presented in figure. \ref{annulus}: there are two triangles but the dot diagram is a little bit unclear from
this perspective. 

The triangulation can be best understood from the disc case using the following trick: one degenerate one of the hole and create two new
marked points which are taken as the full puncture, then one use the familiar triangulations of fourth puncture disc to find the quiver and identify the boundary
quiver nodes on the two edges representing the  full puncture. The BPS quiver should include these identified boundary nodes.
The quiver we find is exactly the same as found in the literature. The underlying network construction has 
several useful advantages, for example,  we know we should mutate quiver nodes $(1,2,3)$ together which correspond to the flip of the triangulation, the new quiver is 
isomorphic to the original quiver. Such quiver mutations are very important for finding the BPS spectrum. Moreover, the potential for this quiver is also manifest from
the network construction.

\begin{figure}[htbp]
\small
\centering
\includegraphics[width=10cm]{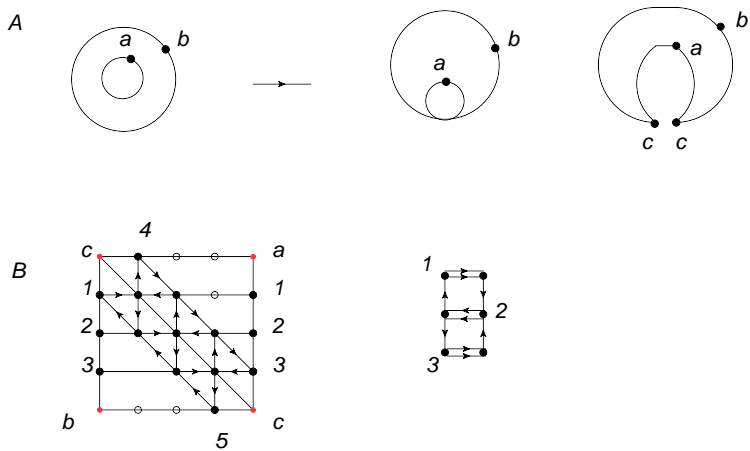}
\caption{ A:  The degeneration limit of the annulus creates two new full punctures. The cyclic order for the four punctures are $a,c,b,c$, where $a,b$ are 
the simple punctures and $c$ is full puncture.
B. We first construct the dot diagram, network(neglected) and the quiver for the four punctured disc. At the end, we need to identify the quiver nodes $
(1,2,3)$. Finally we find the quiver for pure Super Yang-Mills theory. }
\label{pure}
\end{figure}

\begin{figure}[htbp]
\small
\centering
\includegraphics[width=4cm]{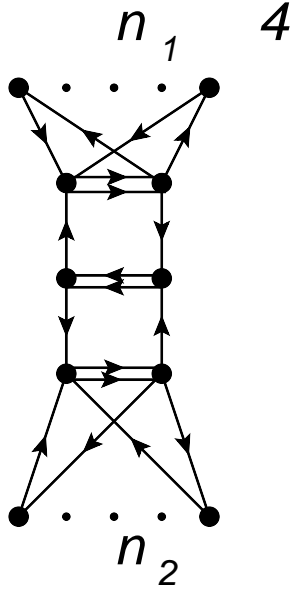}
\caption{We add one more quiver node and a triangle of the quiver arrows to the boundary of the quiver of the pure Yang-Mills theory for each extra fundamentals. }
\label{flavor}
\end{figure}

The irregular singularity for $n_1=1$ is type II singularity with $n=-1$ and $j=1$, and the boundary node for this simple puncture should be 
gauged.  For generic $n_f$, the number of Stokes rays are not enough to describe the moduli space. However, motivated by the above treatment of the $n_f=1$, 
we conjecture that we get $n_f$ extra nodes which are connected to the two nodes of pure $SU(N)$ by a cyclic triangle, see figure. \ref{flavor} . This is 
in agreement with the result in \cite{Alim:2011kw}.

If $n_1=k-1$, then the irregular singularity is the type III singularity and there are two simple punctures on the boundary. If $n_f=0$ on the other boundary, the quiver is found 
by starting with a triangulation of the annulus with two simple punctures and one simple puncture on two boundaries respectively. The quiver is 
best understood from first decomposing the annulus into the disc, and there are two new full punctures which will be get identified in the end.

Notice that there are many choices for the triangulation given a decomposition of the number of flavors $n_f=n_1+n_2$, and $n_1, n_2$ are 
the flavors represented by each boundary, it would be interesting to check that the quiver corresponding to difference choices are in the same mutation class. 
With this construction, we can find the quiver for $SU(k)$ gauge theory with any number of flavors $n_f<k$.

\subsubsection{Linear quiver}
\begin{figure}[htbp]
\small
\centering
\includegraphics[width=4cm]{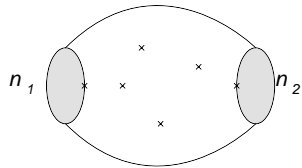}
\caption{ The bordered Riemann surface for the quiver gauge theory $n_1-SU(k)-\ldots-SU(k)-n_2$. All the punctures are the simple, and the special treatment is needed 
on the boundary simple puncture if $n_1,n_2\geq 1$. }
\label{border}
\end{figure}

Now let's consider the generic non-conformal quiver \footnote{This quiver is the $\mathcal{N}=2$ theory, which should not be confused with the BPS quiver.} gauge theory.
The simplest one is the following quiver $n_1-SU(k)-\ldots-SU(k)-n_2$, the six dimensional construction involves two irregular singularities describing $n_1$ and $n_2$ flavors 
for $SU(k)$, there are also several simple punctures in the bulk, the bordered Riemann surface for this theory is shown in figure. \ref{border}. The quiver is found by finding  
the triangulation  and dot diagram of each triangle.

In general, we need to consider the quiver tail represented by the Young Tableaux with rows $r_1, r_2,\ldots, r_s$, and the quiver tail is 
\begin{equation}
SU(h_1)-SU(h_2)-\ldots-SU(h_s)
\end{equation}
where $h_i=\sum_{j=1}^{i}r_j$ and $h_s=k$. The quiver tail we have studied above has a single row with length $k$.
The general linear quiver which is UV complete has the following form
\begin{equation}
Tail_1-SU(k)-SU(k)-\ldots-SU(k)-Tail_2
\end{equation}

The irregular singularity for the quiver tail without any fundamentals has the following form
\[ \Phi=\left( \begin{array}{ccccc}
B_1 & 0 & 0 &0&....\\
0 & B_2 & 0 &0&....\\
0 & 0 & B_3&0&...\\
 0 & 0 & 0&B_4&...\\
 .&.&.&.&...
\end{array} \right)\]
\label{set}
and $B_i={1\over z^{1+{1\over r_i}}}diag(1,\omega,\ldots, \omega^{r_i-1})$ with $\omega=\exp({2\pi i\over r_i})$.  However, generally speaking the number of Stokes rays is 
not enough to describe the full framed moduli space. The special case is if $r_1=r_2=...=r_s$ and the irregular singularity is the type I singularity with $n=-1$ and $j={k\over r_s}$.
There are $j$ simple punctures on the boundary corresponding to this irregular singularity.  After that, the quiver and the triangulation is straightforward to find.

\textbf{Example}: Consider the quiver $SU(2)-SU(4)-SU(6)$, and the Young Tableaux has rows $2,2,2$. There are three simple punctures on one boundary and one simple 
puncture on the other boundary.  The number of charge lattice of this theory is $20$ which is the same as the quiver nodes from the triangulation of the bordered Riemann surface.

Another special case is the quiver tail 
\begin{equation}
1-SU(2)-SU(3)-\ldots-SU(k)
\end{equation}
the irregular singularity is a  type I singularity with $n=0$ and $j=0$ and the leading order coefficient is regular semi-simple, so there are two full punctures on the boundary.
The BPS quiver for the theory with this tail is then easy to find. It would be interesting to have other method to find the BPS quiver of general quiver tail.

\subsection{Theory with AD matter}
Now let's consider the asymptotical free theory defined by a Riemann surface with arbitrary number of irregular and regular singularities. The 
$\mathcal{N}=2$ theory is a quiver gauge theory coupled with type IV AD theory and isolated theory represented by three regular singularities on a sphere.

To find the BPS quiver for this theory, one first blow up the irregular singularity into a disc with labeled marked points, and we have a 
bordered Riemann surface. Then we can find the triangulations, network and the quiver. For genus zero case, there is another construction coming from 
the disc: one first degenerate the boarded Riemann surface into a disc,  and there are two new full punctures \footnote{This fact is similar to the theory of class ${\cal S}$ in which
two full punctures appear if a handle is removed, the difference is that there the punctures appear in the bulk of the Riemann surface.} appearing every time a hole is removed,
and these new punctures are located at the boundary of the disc. Notice that there is a hidden cyclic order for all the punctures on the disc.
The construction of the network of the disc with marked points is well understood, and finally we identify the quiver nodes associated with those 
punctures appearing from degenerating from the hole.

\begin{figure}[htbp]
\small
\centering
\includegraphics[width=10cm]{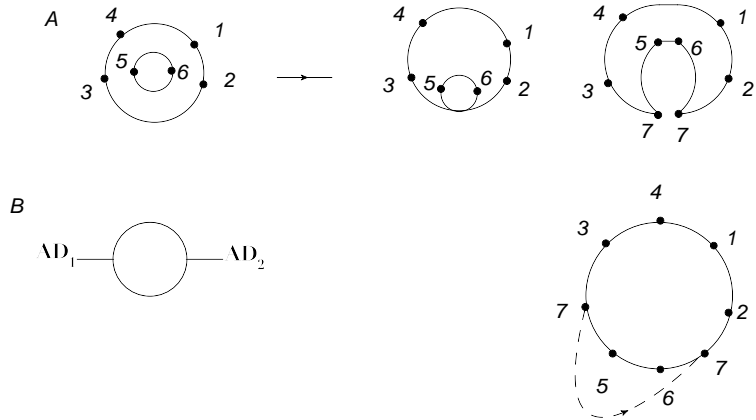}
\caption{This is a $SU(k)$ theory coupled with two AD theories, all the marked points are full. The quiver can be found from the disc with marked points and then glue two punctures coming 
from the degeneration. }
\label{AF1}
\end{figure}

\textbf{Example} This theory has two irregular singularities which has integer order with leading order regular semi-simple, so the boarded Riemann surface 
model has two boundaries with full marked points. The gauge theory is a $SU(k)$ gauge group coupled with two Type IV
theories which are realized as the Riemann sphere with one irregular singularity and one full regular singularity. The disc after the degeneration 
is shown in figure. \ref{AF1}, 

\newpage
\section{Conclusion}
We construct  quiver and cluster coordinate for a large class of AD theory and asymptotical theory in this paper using the Stokes data for the irregular singularity. 
The planar network for type I  or type II AD theory actually describes a positive cell in $G(.,k)$ Grassmannia. The constructed network 
is automatically minimal and we hope our result could be useful for the study of the scattering amplitude in which such network plays
a very important role \cite{Nima:2012}. These planar networks are also used in constructing four dimensional $\mathcal{N}=1$ SCFT \cite{Xie:2012mr,Franco:2012mm} and our construction  
leads to so-called minimal duality frame. It is amazing that such simple combinatorial objects appear in so many completely different 
physical systems, it is definitely interesting to find the deep connections. 

The quiver and quiver mutation are important in the study of the cluster coordinates. However, it is very hard to prove that two quivers are related by quiver mutations,  especially if the quiver mutation class is 
infinite, which is the general case considered in this paper. The remarkable thing of the planar network is that one can actually define some 
quiver mutation invariant quantity using simple combinatorial tools. In fact, one can define a permutation  \cite{postnikov-2006} which uniquely fixed the network class
related by square moves, therefore we can judge whether two quivers from networks are in the same mutation class.
For our purpose, the network has some redundant information since we only 
care about the closed surface, so two quivers can be in the same mutation class even if the two networks are not in the same class. The permutation 
method is only a necessary condition but not the sufficient condition for the quiver mutation class.  Moreover, there are networks defined on non-planar surface and
no general method is known to judge whether two networks are related by square move. This question is deserved further study.

It would be interesting to compare our result with the spectral network construction proposed in \cite{Gaiotto:2012rg}, one possible clue is that the 
network can be equivalently replaced by a triple point diagram which seems to be closed related to some special configurations of spectral network.

We have not  studied the cluster coordinates in any detail in this paper, such coordinates are important for studying the BPS spectrum and 
wall crossing behavior \cite{Xie:2012ww}, in particular, quiver mutation sequences which lead back to original quiver can be easily found from
our construction. Moreover, the network construction provides a very natural potential for the quiver, and the representation theory of quiver with potential
could be used to study the spectrum and wall crossing. 
They are also very useful in studying the underlying hyperkahler metric of the Hitchin's moduli space \cite{Gaiotto:2008cd, Xie:2012hh}.

The line operator is related to the perfect matching of the network and one can define a boundary measurement using the cluster 
coordinates associated with the surfaces \cite{postnikov-2006}. The boundary measurement is invariant under the square move which could be thought 
of as certain invariant information on crossing the wall, it is conceivable that these boundary measurements give the expectation value of the line operators.
we believe many important physical information could be extracted from this fact, i.e. the framed BPS states counting.

A cluster integrable system \cite{goncharov-2011} can be defined on the network which could be related to the Seiberg-Witten integrable system, and 
the quantization using cluster coordinates is relatively easy. There is a very nice
Poisson structure defined using the quiver and the quantization of the cluster coordinates should be related to the Nekrasov
partition function. It is interesting to carry out this calculation in detail. 

The cluster coordinate transformation for some special quiver mutations defines certain $Y$ system \cite{Fomin_y-systemsand}, which is important for studying scattering amplitude and form factor
of $\mathcal{N}=4$ super Yang-mills theory in strongly coupling limit \cite{Alday:2009dv, Maldacena:2010kp}. We have explicitly constructed such $Y$ system for a large class of cases, hopefully, this result 
could be useful in that context.

\begin{flushleft}
\textbf{Acknowledgments}
\end{flushleft}
This research is supported in part by Zurich Financial services membership and by the U.S. Department of Energy, grant DE- FG02-90ER40542 (DX).

\bibliographystyle{utphys} 
 \bibliography{PLforRS}

\end{document}